\documentclass{article} 
\usepackage{iclr2015,times}
\usepackage{hyperref}
\usepackage{url}
\usepackage{fixltx2e}
\usepackage{amsmath,graphicx}
\usepackage{amsmath}
\usepackage{amssymb}

\usepackage{bm}

\newcommand{\RR}{\mathbb{R}}

\newcommand{\mypara}[1]{\smallskip\noindent\textbf{#1\,}}

\title{Audio Source Separation with Discriminative Scattering Networks}

\author{
Pablo Sprechmann$^1$, Joan Bruna$^2$, Yann Lecun$^{1,2}$ \\
$^1$ NYU, Courant Institute of Mathematical Sciences, $^2$ Facebook AI Research. \\
\texttt{ \{pablo,bruna,yann\}@cims.nyu.edu} \\
}

\iclrfinalcopy 

\begin{document}

\maketitle

\begin{abstract}
In this report we describe an ongoing line of research
for solving single-channel source separation problems.
Many monaural signal decomposition techniques proposed in the literature operate on a feature space consisting of a time-frequency representation of the input data.
A challenge faced by these approaches is to effectively exploit the temporal dependencies of the signals at scales larger than the duration of a time-frame.
In this work we propose to tackle this problem by modeling the signals using a time-frequency representation with multiple temporal resolutions.
The proposed representation consists of a pyramid of wavelet scattering operators, which generalizes Constant Q Transforms (CQT) 
with extra layers of convolution and complex modulus. 
We first show that learning standard models with this multi-resolution setting improves
source separation results over fixed-resolution methods. 
As study case, we use Non-Negative Matrix Factorizations (NMF) that has been widely considered in many audio application. 
Then, we investigate the inclusion of the proposed multi-resolution setting into a discriminative training regime. 
We discuss several alternatives using different deep neural network architectures. 
\end{abstract}

$ $
\section{Introduction}
\label{sec:intro}

Monaural Source Separation is a fundamental inverse problem in 
speech processing (\citet{loizou2007speech,hansler2008speech}). 
Successful algorithms rely on models that capture signal regularity 
while preserving discrimination between different speakers.  
The decomposition of time-frequency representations, such as the power or magnitude spectrogram
in terms of elementary atoms of a dictionary, has become a popularcitet tool in audio processing. 
Non-negative matrix factorization (NMF) (\citet{NMF}),
have been widely adopted in various audio processing tasks, including in particular source separation, see \citet{smaragdis2014static} for a recent review. 
There are many works that follow this line in speech separation (\citet{schmidt06speechseparation,shashanka_icassp07}) and enhancement (\citet{DuanMS12,mohammadiha2013supervised}). 

Although NMF applied on spectral features is highly efficient, it fails to model 
long range geometrical features that characterize speech signals. Increasing the 
temporal window is not the solution, since it increases significantly the dimensionality of the problem and reduces the discriminative power of the model.
In order to overcome this limitation,  
many works have proposed regularized extensions of NMF to promote learned structure in the codes. 
Examples of these approaches are, temporal smoothness of the activation coefficients (\citet{fevotte2011majorization}), 
including  co-occurrence statistics of the basis functions (\citet{WilsonRSD08}), and learned temporal dynamics with Kalman filtering
like techniques(\citet{MysoreS11,HanMP12,icassp13a}) or integrating Recurrent Neural Networks (RNN) into the NMF framework (\citet{BL}).

More recently, several works have observed that the efficiency of these methods can be improved with discriminative training. 
Discriminatively trained dictionary learning techniques (\citet{mairal2012task, LecunNN,sprechmann2014supervised, weninger2014discriminative}) show the importance 
of adapting the modeling task to become discriminative at the inverse problem at hand. A number of works completely bypass the modeling 
aspect and approach inverse problems 
as non-linear regression problems using Deep Neural Networks(DNN)  (\citet{sprechmann2013learnable, deblur_mpi, superres}) with differet levels of structure ranging from simple frame-by-frame regressors
to more sophisticated RNN. Applications include source separation in music (\citet{sprechmann2012real,huang2014singing}), speech separation (\citet{Huang_DNN_Separation_ICASSP2014}) and speech enhancement (\citet{Weninger2014GlobalSIP12}).

The goal of this work is to show that using stable and robust multi-resolution representation of the data
can benefit the sources separation algorithms in both discriminative and non-discriminative settings.
Previous works have shown that the choice of the input features plays a very important role of
on source separation (\citet{Weninger2014GlobalSIP12}) and speech recognition (\citet{mohamed2012understanding}).
This work takes this observation a step further to the multi-resolution setting.

We consider a deep representation based on the wavelet scattering pyramid, which produces information at different temporal resolutions and
defines a metric which is increasingly contracting. This representation can be thought
as a generalization of the CQT.
Discriminative features having longer temporal context can be constructed with the 
scattering transform (\citet{pami}) and have been sucessfully applied
to audio signals by \citet{deepscatt}. 
While these features have shown excellent performance in various classification tasks, 
in the context of source separation we require a representation that not only captures
long-range temporal structures, but also preserves as much temporal discriminability as possible.

For the non-discriminative setting, we present an extension of the NMF framework to the pyramid representation. 
We learn NMF models at different levels
of the hierarchy. While NMF dictionaries at the first level are very selective to temporally localized energy 
patterns, deeper layers provide additional modeling of the longer temporal dynamics (\citet{icassp14}).
For the discriminative setting we discuss a number of baseline models based on neural networks. 
%
As a proof of concept, we evaluate both settings on a multi-speaker speech separation task. 
We observe that in both training regimes the multi-resolution setting leads to better performance
with respect to the baselines. 
We also confirm with experiments the superiority of discriminative approaches.


The paper is organized as follows.
In Section~\ref{sec:sss}  we describe the general setting of source separation and review
some baseline solutions for both in training regimes. We present the proposed representation in Sections~\ref{sec:pyr} and show how it can be used
in the context of source separation in Section~\ref{sec:training}. We show some initial experimental results in Section~\ref{sec:exp}
and a discussion is given in Section~\ref{sec:disc}.

\section{Single-channel source separation}
\label{sec:sss}

In this work we are intereseted in the families of algorithms that solve source separation
on a feature space. This section is dedicated to describing different alternatives that fall in this category.
We first introduce the general setting in Section~\ref{sec:problem.from}.
In Section~\ref{sec:nmf} we describe the popular NMF framework 
and different training regimes employed with it. Finally we discuss
purely discriminative approaches based on deep networks in Section~\ref{sec:deep}.

\subsection{Problem formulation}
\label{sec:problem.from}

We consider the setting in which we observe a temporal signal $y(t)$ that is the sum of  
two sources $x_i(t)$, with $i=1,2$,
\begin{equation}
\label{ssep}
y(t) = x_1(t) + x_2(t),
\end{equation}
and we aim at finding estimates $\hat{x_i}(t)$. 
We consider the supervised 
monoaural source separation problem, in which the components $x_i$, $i=1,2$ come from sources for 
 which we have representative training data. 
 In this report we concentrate to the case of speech signals, but other alternatives could be considered, such as noise or music.

Most recent techniques typically operate on a non-negative time-frequency representation.
Let us denote as $\Phi(y) \in \RR^{m \times n}$ the transformed version of $y(t)$, comprising $m$ frequency bins and $n$ temporal frames. 
This transform can be thought as a non-linear analysis operator and is typically defined as the magnitude (or power) of a time-frequency representation such
as the Short-Time Fourier Transform (STFT). Other robust alternatives have also been explored (\citet{Huang_DNN_Separation_ICASSP2014,Weninger2014GlobalSIP12}).
In all cases, the temporal resolution of the features is fixed and given by the frame duration. 

Performing the separation in the non-linear representation is key to the success of these algorithms. The
transformed domain is in general invariant to some irrelevant variability of the signals (such as local shifts), thus relieving the 
algorithms from learning it. This comes at the expense of inverting the unmixed estimates in the feature space, normally known as the
phase recovery problem (\citet{yonina}). 
Specifically, these algorithms take $\Phi(y)$ as input and produce estimates for each source, $\Phi(\hat{x}_i)$ with $i=1,2$. 
The phase recovery problem corresponts to finding signals $\hat{x}'_i$ such matching the obtained features $\Phi(\hat{x}_i)$
and satisfying $y=\hat{x}'_1 + \hat{x}'_2$.

The most common choice is to use the magintud (or power) STFT as the feature space. In this case, the phase recovery problem can be solved very efficiently using 
soft masks to filter the mixture signal (\citet{schmidt07mlsp}). The strategy resembles Wiener filtering and has demonstrated very good results in practice.
Specifically, $\Phi(y) = | \mathcal{S} \{y \}|$, where $\mathcal{S} \{y \}  \in \mathbb{C}^{m \times n}$ is a complex matrix corresponding to the STFT.
The estimated unmixed signals are obtained by filtering the mixture, 
\begin{equation}
\hat{x}_i = \mathcal{S}^{-1} \left\{  M_i\circ \mathcal{S} \{y\}\right \}, \quad \textrm{with} \quad M_i =\frac{\Phi(\hat{x}_i)^p }{ \sum_{l=1,2} \Phi(\hat{x}_l)^p},
\label{eq:rec}
\end{equation}
where multiplication denoted $\circ$, division, and exponentials are element-wise operations. The parameter $p$ defines the smoothness of the mask, 
we use $p=2$ in our experiments. Note that this solution automatically imposes the consistency restriction
$y=\hat{x}'_1 + \hat{x}'_2$.

\subsection{Non-negative matrix factorization}
\label{sec:nmf}

Source separation methods based on matrix factorization approaches have received a lot of attention in the literature in recent years.
NMF-based source separation techniques attempt to find the non-negative activations $Z_i \in \RR^{q \times n}$, $i=1,2$ 
best representing the different speech components in two dictionaries ${D}_{i} \in \RR^{m \times q}$.
Ideally one would want to solve the problem,
\begin{equation}
\label{ideal_model}
\min_{x'_i, Z_i\geq 0} \sum_{i=1,2} \mathcal{D}( \Phi(x'_i) | D_i Z_i ) + \lambda \mathcal{R}(Z_i)\quad\,s.t. ~y=x'_1 + x'_2~.
\end{equation}
where the first term in the optimization objective measures the dissimilarity between the input data and the estimated channels in
the feature space. 
Common choices of $\mathcal{D}$ are the squared Euclidean distance,
the Kullback-Leibler divergence, and the Itakura-Saito divergence. The second term in the minimization objective is included to promote some desired structure of the activations. 
This is done using a designed regularization function $\mathcal{R}$, whose relative importance is controlled by the parameters $\lambda$. 
%
In this work we use $\mathcal{D}$ reweighted squared Euclidean distance and the $\ell_1$ norm as the regularization function $\mathcal{R}$. 

Problem (\ref{ideal_model}) could be minimized with an alternating gradient descent between $x'_i$ and $z_i$. 
Note that fixing $z_i$ and minimizing with respect to $x'_i$ requires locally inverting the transform 
 $\Phi$, which amounts to solve an overcomplete phase recovery problem. 
In practice, a greedy proxy of (\ref{ideal_model}) is solved instead. First a separation is obtained in the feature spaces by solving
a clasic NMF problem,
\begin{eqnarray}
\label{eq:optim_general}
\min_{ Z_i \ge 0 } D( \Phi(x) |  \, \sum_{i=1,2} {D}_i Z_i  ) + 
\lambda\, \sum_{i=1,2} \mathcal{R}(Z_i)~,
\end{eqnarray}
for which there exist standard optimization algorithms, see for example \citet{fevotte2011algorithms}. Once the optimal activations are solved for, the spectral envelopes of the speech are estimated as $\Phi(\hat{x_i}) = {D}_{i} Z_{i}$, and the phase recovery is solved using (\ref{eq:rec}).

In this supervised setting, the dictionaries are obtained from training data. The classic approach is to build model for each source independently and
later use them together at testing time.
Many works have observed that sparse coding inference algorithms can be improved in specific tasks by using discriminative training, i.e. 
by directly optimizing the parameters of the model on the evaluation cost function. Task-aware (or discriminative) sparse modeling 
is elegantly described by \citet{mairal2012task}, observing that one can back-propagate through the Lasso.  
These ideas have been used in the context of source separation and enhancement (\citet{sprechmann2014supervised, weninger2014discriminative}).
The goal is to obtain dictionaries such that the solution of (\ref{eq:optim_general}) also minimizes the reconstruction given the ground
truth separation,
\begin{eqnarray}
\min_{D_1 \ge 0 , D_2 \ge 0 } \,\,\mathcal{D}( \Phi(x_1) |  {D}_1  Z_1^\ast  )
 + \alpha \mathcal{D}( \Phi(x_2) |  {D}_2  Z_2^\ast ),
\label{eq:bilevel}
 \end{eqnarray}
where $Z_i^\ast$  are the solutions of (\ref{eq:optim_general}) (and depend on the dictionaries) and $\alpha$ is a parameter controlling the relative importance of source recovery; typically, one would set $\alpha=0$ in a denoising application (where the second signal is noise), and $\alpha = 1$ in a source separation application
where both signals need to be recovered. When the phase recovery can be obtained using the masking scheme described in Section~\ref{sec:problem.from}, it could
be included into the objective in order to directly optimize the signal reconstruction in the time domain.
While the discriminative setting is a better target, the estimation needs to be computed over the product set rather than each training set independently and the generalization might be compromised when small training sets are available.
It is important to note that the level of supervision is very mild, as in the training of autoencoders. We are artificially generating
the mixtures, and consequently obtaining the ground truth.

The standard NMF approaches treat different time-frames independently, ignoring the 
temporal dynamics of the signals. As described in Section~\ref{sec:intro}, many works attempt to change the regularization function 
$\mathcal{R}$ in order integrate several frames into de decomposition. It's analysis and description is outside the scope of this report.

\subsection{Purely discriminative settings}
\label{sec:deep}

With the mindset of the discriminative learning, one is tempted to simply replace the inference step by a generic neural network architecture, having enough capacity 
to perform non-linear regression. The systems are trained as to minimize a measure of fitness between the ground truth separation and the output as in (\ref{eq:bilevel}),
being the most common the Mean Squared Error (MSE). Not that this can be performed in the feature space or in the time domain (when the phase recovery is simple). Other alternatives studied in the literature consist of predicting
the masks given in (\ref{ideal_model}) as described by \citet{Huang_DNN_Separation_ICASSP2014}.

 The most straight forward choice is to perform the estimation using a DNN
on a fixed time scale. 
Using a short temporal context fails to model long range temporal dependencies on the speech signals, while increasing the context
renders the regression problem intractable. One could consider to train a DNN on an set of several frames (\citet{Weninger2014GlobalSIP12}).
Recent works have explored neural network architectures that exploit
temporal context such as RNN and Long Short-Term Memory (LSTM) (\citet{Huang_DNN_Separation_ICASSP2014,Weninger2014GlobalSIP12}).
\section{Pyramid Wavelet Scattering}
\label{sec:pyr}

In this section we present briefly the proposed wavelet scattering pyramid,
which is conceptually similar to standard scattering networks introduced by \citet{eurispco}, 
but creates features at different temporal resolutions at every layer.

\subsection{Wavelet Filter Bank}

A wavelet $\psi (t)$ is a band-pass filter with good frequency and spatial localization.
 We consider a complex wavelet with a quadrature phase, 
whose Fourier transform satisfies
$\mathcal{F} \psi(\omega) \approx 0$ for $\omega < 0$.
We assume that the center frequency of $\mathcal{F} \psi$ is $1$ and 
that its bandwidth is of the order of $Q^{-1}$. 
Wavelet filters centered
at the frequencies $\lambda = 2^{j/Q}$ are computed by dilating $\psi$:
$\psi_\lambda (t) = \lambda\, \psi(\lambda\, t)$, and hence $\mathcal{F} \psi_\lambda (\omega) = \widehat \psi(\lambda^{-1} \omega)$.
We denote by $\Lambda$ the index set of $\lambda = 2^{j/Q}$ over
the signal frequency support, with $j \leq J_1$. 
The resulting filter bank has a constant number $Q$ of bands per 
octave and $J_1$ octaves. Let us define $\phi_1(t)$ as a low-pass filter
with bandwidth $~2^{-J_1}$.
The wavelet transform of a signal $x(t)$ is
\[
W x = \{x \ast \phi_1(t)~,~x \ast \psi_\lambda(t)  \}_{\lambda \in \Lambda}~.
\]
Since the bandwidth of all filters is at most $Q^{1}$, we can down-sample 
its outputs with a stride $~Q$.

\subsection{Pyramid Scattering Transform}

Instead of using a fixed bandwidth smoothing kernel that is applied 
at all layers, we sample at critical rate in order to preserve temporal locality 
as much as possible. 
We start by removing the complex phase of wavelet coefficients in $Wx$ with a 
complex modulus nonlinearity.
Then, we arrange these first layer coefficients as nodes in the first level of a tree.
 Each node of this tree is down sampled at the critical sampling rate 
of the layer $\Delta_1$, given by the reciprocal of the largest bandwidth present
in the filter bank:
\[
|W^1| x = \{ x^1_i \}_{i=1\dots 1+|\Lambda|}= \{x \ast \phi_1(\Delta_1 n)~,~|x \ast \psi_\lambda(\Delta_1 n)|  \}_{\lambda \in \Lambda}~.
\]

These first layer coefficients give localized information both in time and frequency, 
with a trade-off dictated by the $Q$ factor. They are however sensitive to local 
time-frequency warps, which are often uninformative. In order to increase the 
robustness of the representation, we transform each of the down sampled signals 
with a new wavelet filter bank and take the complex modulus of the oscillatory component. 
For simplicity, we assume a dyadic transformation, 
which reduces the filter bank to a pair of conjugate mirror filters $\{ \phi_2, \psi_2\} $ (\citet{wavelettour}), 
carrying respectively the low-frequencies and high-frequencies of the discrete signal from above the tree:
$$|W^2| x = \{ x^1_i \ast \phi_2 (2 n)~,~| x^1_i \ast \psi_2 (2n)| \}_{i=1\dots |W^1|}~.$$
Every layer thus produces new feature maps at a lower temporal resolution. As
shown in \citet{pami}, only coefficients having gone through $m \leq m_{max}$ non-linearities 
are in practice computed, since their energy quickly decays. We fix $m_{max}=2$ in our experiments. 

We can reapply the same operator as many times $k$ as desired until reaching a temporal 
context $T = 2^k \Delta_1$. If the wavelet filters are chosen such that they define a non-expansive mapping \citet{pami}, 
it results that every layer defines a metric which is increasingly contracting:

$$\| |W^k| x - |W^k| x' \| \leq \| |W^{k-1}| x - |W^{k-1}| x' \| \leq \| x - x' \| ~.$$
Every layer thus produces new feature maps at a lower temporal resolution. 
In the end we obtain a tree of different representations, $\Phi_j(x) = |W^j| x$ with $j=1,\ldots, k$.

\section{Source Separation Algorithms}
\label{sec:training}

In this section we show a few examples of how the proposed pyramid scattering features
could be used for solving the source separation problem. 
We present alternatives for both learning paradigms: non-discriminative and discriminative. 

\subsection{Non-Discriminative Training}
\label{sec.non-disc}

In this setting, we try to find models for each speaker using the features of the 
wavelet scattering pyramid.
Each layer of the transform produces information with different stability/discriminability trade-offs. 
Whereas in typical classification applications one is mostly interested in choosing a single layer which provides 
the best trade-off given the intrinsic variability of the dataset, in inverse problems we can leverage signal models
at all levels. Let us suppose two different sources $X_1$ and $X_2$, and let us consider for simplicity the features
$\Phi^j(x_i)$, $j=1,2$, $i=1,2$, $x_i \in X_i$, obtained by localizing the scattering features of two different resolutions at their 
corresponding sampling rates. Therefore, $\Phi_1$ carries more discriminative and localized information than $\Phi_2$. 

In the non-discriminative training, we train independent models for each source. 
Given training examples $X^t_i$ from each source, we consider a NMF of each of the features $\Phi_j (x^t_i)$:
$$\min_{D^j_i , Z^j_i \geq 0}  \,\, \sum_{x_i \in X_i}\, \frac{1}{2}  \| \Phi^j(x_i) - D^j_i Z^j_i \|^2 + \lambda^{j}_i \| Z^j_i \|_1,$$
where here the parameters $\lambda^{j}_i$ control the sparsity-reconstruction trade-off in the sparse coding. In our experiments we used a fixed value for
all of them $\lambda^{2}_i=\lambda$. At test time, given $y=x_1 + x_2$, we estimate $\hat{x}_1$, $\hat{x}_2$ as the solution of
\begin{equation}
\label{info}
\min_{\hat{x}_1 + \hat{x}_2 = y, Z^j_i\geq 0} \sum_{i=1,2} \frac{1}{2} \| \Phi^1(\hat{x}_i) - D^{1}_i Z^{1}_i \|_2^2 + \lambda^{1}_i \| Z_i \|_1
+ \frac{1}{2} \| \Phi^2(\hat{x}_i) - D^{2}_iZ^{2}_i \|_2^2 +\lambda^{2}_i \| Z^{2}_i \|_1 ~.
\end{equation}
Problem (\ref{info}) is a coupled phase recovery problem under linear constraints. It can be solved using gradient descent as in \citet{icassp_sounds}, 
but in our setting we use a greedy algorithm, which approximates the unknown complex phases using the phase of $W_1 y$ and $W_2 | W_1 y| $ respectively. 
Similarly as in \citet{Weninger2014GlobalSIP12}, we simplify the inference by using a stronger version of the linear constraint $y=x_1 + x_2$, namely 
$$|W^1 y|^2 = |W^1 x_1|^2 + |W^1 x_2|^2~,$$ 
and therefore that destructive interferences are negligible. 

\subsection{Discriminative Training}
\label{sec.disc}

The pyramid scattering features can also be used to train end-to-end models.
The most simple alternative is to train a DNN directly from features having the same temporal context as second layer scattering features. 
For simplicity, we replace the second layer of complex wavelets and modulus with a simple Haar transform:
$$\Phi_2(x) = \{ | x \ast \psi_\lambda | \ast h_k (\Delta_1 n) \}_{\lambda \in \Lambda, k=0,\dots J_2}~,$$
where $h_k$ is the Haar wavelet at scale $2^ k$, and we feed this feature into a DNN with the same number of hidden units as before. 
We do not take the absolute value as in standard scattering to leave the chance to the DNN to recombine coefficients before the first non-linearity. 
We report results for $J_2=5$ which corresponds to a temporal context of $130$ms. We will refer to this alternative as \emph{DNN-multi}. 
As a second example, we also consider a multi-resolution Convolutional Neural Network (CNN), constructed by creating contexts of three temporal frames at resolutions $2^j$, $j=0\dots, J_2=5$. We will refer to this alternative as \emph{CNN-multi}. This setting has the same temporal context as the \emph{DNN-multi} but rather than imposing separable filters we leave extra freedom. This architecture can access relatively large temporal context with a small number of learnable parameters. 
Since the phse recovery problem cannot be approximated with softmax as in (\ref{ideal_model}), we use as the cost function the MSE of the reconstructed feature at
all resolutions.
\section{Experiments}
\label{sec:exp}

In this section we present some initial experimental evaluation in which we study
the use of multi resolution signal representation with both discriminative and non-discriminative training regimes. 
We compare the performance against some basic baseline settings. 

As a proof of concept, we evaluated the different alternatives in a multi-speaker setting in which we aim at separating
male and female speech. In each case, we trained two gender-specific modeles. The training data consists of recordings of a generic group of speakres per gender, none of which were included in the test set.
%
The experiments were carried out on the TIMIT corpus. 
We adopted the standard test-train division, using all the training recordings (containing 462 different speakers) for building the models
and a subset of 12 different speakers (6 males and 6 females) for testing. For each speaker we randomly chose two clips and compared
all female-male combinations (144 mixtures).  
All signals where mixed at 0 $dB$ and resampled to $16$ kHz. 
We used the \emph{source-to-distortion ratio} (SDR), \emph{source-to-interference ratio} (SIR), and
\emph{source-to-artifact ratio} (SAR) from the BSS-EVAL metrics (\citet{vincent2006performance}). We report the average
over the both speakers, as the measure are not symmetric.

\mypara{Non-discriminative settings: }  As a basline for the non-discriminative setting we used  standard NMF 
with STFT of frame lengths of 1024 samples and 50\% overlap, leading to $513$ feature vectors. The dictionaries were chosen with $200$ and $400$ atoms. We evaluated the proposed scattering features in combination with NMF (as described in Section~\ref{sec.non-disc}) with one and two layers,  referred
as \emph{scatt-NMF\textsubscript{1}} and \emph{scatt-NMF\textsubscript{2}} respectively. 
We use complex Morlet wavelets with $Q_1=32$ voices per octave in the first level, and dyadic Morlet wavelets ($Q_2=1$) for the second level,
for a review on Morlet wavelets refer to \citet{wavelettour}.
The resulting representation had $175$ coefficients for the first level and around $2000$ for the second layer. 
%
We used $400$ atoms for \emph{scatt-NMF\textsubscript{1}}  and $1000$ atoms for \emph{scatt-NMF\textsubscript{2}}.
In all cases, the features were frame-wise normalized and we used $\lambda=0.1$. 
In all cases, parameters were obtained using cross-validation on a few clips separated from the training as a validation set. 

\mypara{Discriminative settings: } We use a single and multi-frame \emph{DNN}s as a baseline for this training setting.The network architectures consist of two hidden layers using the outputs of the first layer of scattering, that is, the CQT coefficients at a given temporal position. It uses RELU's as in the rest of the architectures and the output is normalize so that it corresponds to the spectral mask discussed in (\ref{ideal_model}).
The multi-frame version considers the concatenation of 5 frames as inputs matching the temporal
 context of the tested multi-resolution versions. We used $512$ and $150$ units for the single-frame DNN (referred as \emph{CQT-DNN}) and $1024$ and $512$ for the multi-frame one (referred as \emph{CQT-DNN-5}), increasing the number of parameters did not improve the results.
We optimize the network to optimize the MSE to each of the sources. We also include the architectures \emph{DNN-multi} and
\emph{CNN-multi} described in Section~\ref{sec.disc}.
In all cases the weights are randomly initialized and training is 
performed using stochastic gradient descent with momentum. We used the GPU-enabled package Matconvnet (\citet{matconvnet}).

\begin{table}[t]
\centering
\begin{tabular}{l|c|c|c }
  \hline\hline
 & SDR & SIR & SAR\\
\hline
NMF  & 6.1 [2.9] &   14.1 [3.8] & 7.4 [2.1] \\
\hline
\emph{scatt-NMF\textsubscript{1}} &   6.2 [2.8] &   13.5 [3.5] & 7.8 [2.2] \\
\emph{scatt-NMF\textsubscript{2}} &    6.9 [2.7] &  16.0 [3.5]  &  7.9 [2.2] \\
\emph{CQT-DNN} &   9.4 [3.0] &  17.7 [4.2]  &  10.4 [2.6]   \\
\emph{CQT-DNN-5}  &9.2 [2.8] &  17.4 [4.0]  &  10.3 [2.4]   \\
\emph{CQT-DNN-multi} &   9.7 [3.0] &  19.6 [4.4]  & 10.4 [2.7]   \\
\emph{CQT-CNN-multi} &   {\bf 9.9} [3.1] &  {\bf19.8} [4.2]  & {\bf 10.6} [2.8]   \\
  \hline
  \hline
\end{tabular}
\caption{Source separation results on a multi-speaker settings. Average SDR, SIR and SAR (in $dB$) for different methods. Standard deviation of each result shown between brackets. 
\label{ta:eval}}
\vspace{-2ex}
\end{table}

Table~\ref{ta:eval} shows the results obtained for the speaker-specific and multi-speaker settings. 
In all cases we observe that the one layer scattering transform outperforms the STFT in terms of SDR.
Furthermore, there is a tangible gain in including a deeper representation; \emph{scatt-NMF\textsubscript{2}} 
performs always better than \emph{scatt-NMF\textsubscript{1}}. While the gain in the SDR and SAR are relatively small the SIR is 3dB higher.
It is thus benefitial to consider a longer temporal context in order to perform the separation
sucessfully.

On the other hand, as expected, the discriminative training yields very significant improvements. The same reasons that produced the improvements 
in the non-discriminative setting also have an impact in the discriminative case. Adding enough temporal contexts to the neural regressors improves 
their performance. The multi-temporal representation plays a key role as simply augmenting the number of frames does not lead to better performance (at least using baseline DNNs).
It remains to be seen how these architectures would compare with the alternative RNN models.

\section{Discussion}
\label{sec:disc}
We have observed that the performance of baseline source separation algorithms
can be improved by using a temporal multi-resolution representation.
The representation is able to integrate information across longer temporal contexts while removing uninformative variability 
with a relatively low parameter budget. 
In line with recent findings in the literature, we have observed that including discriminative
criteria in the training leads to significant improvements in the source separation performance.
However, contrary to standard sparse modeling in which the resulting inference can be readily approximated
with a neural network, it remains unclear whether phase-recovery type inference can also be 
efficiently approximated with neural network architectures. We believe 
there might still be a gap in performance that might be bridged with appropriate discriminative architectures.

While this report presents shows some promising initial results, several interesting comparisons need to be made and are subject of current research. 
We consider an interesting problem exploring the best way of including the long-term temporal consistency
into the estimation. Recent studies have evaluate the use of deep RNN's for 
solving the source separation problem \cite{Huang_DNN_Separation_ICASSP2014,Weninger2014GlobalSIP12}.
While \cite{Huang_DNN_Separation_ICASSP2014} do not observe significant improvements
over standard DNN's in speech separation, \cite{Weninger2014GlobalSIP12} obtain significant
improvements using LSTM-DRNN in speech enhancement. We are currently addressing the question of comparing different neural network architectures that exploit
temporal dependancies and assessing whether the use of multi-resolution representation can play a role as in this initial study.

\bibliography{iclr2015_bss_final}

\begin{thebibliography}{37}
\providecommand{\natexlab}[1]{#1}
\providecommand{\url}[1]{\texttt{#1}}
\expandafter\ifx\csname urlstyle\endcsname\relax
  \providecommand{\doi}[1]{doi: #1}\else
  \providecommand{\doi}{doi: \begingroup \urlstyle{rm}\Url}\fi

\bibitem[And{\'e}n \& Mallat(2013)And{\'e}n and Mallat]{deepscatt}
And{\'e}n, J. and Mallat, S.
\newblock Deep scattering spectrum.
\newblock \emph{arXiv preprint arXiv:1304.6763}, 2013.

\bibitem[Boulanger-Lewandowski et~al.(2014)Boulanger-Lewandowski, Mysore, and
  Hoffman]{BL}
Boulanger-Lewandowski, N., Mysore, G.J., and Hoffman, M.
\newblock Exploiting long-term temporal dependencies in nmf using recurrent
  neural networks with application to source separation.
\newblock In \emph{ICASSP}, pp.\  6969--6973, May 2014.

\bibitem[Bruna \& Mallat(2013{\natexlab{a}})Bruna and Mallat]{icassp_sounds}
Bruna, J. and Mallat, S.
\newblock Audio texture synthesis with scattering moments.
\newblock \emph{arXiv preprint arXiv:1311.0407}, 2013{\natexlab{a}}.

\bibitem[Bruna \& Mallat(2013{\natexlab{b}})Bruna and Mallat]{pami}
Bruna, J. and Mallat, S.
\newblock Invariant scattering convolution networks.
\newblock \emph{Pattern Analysis and Machine Intelligence, IEEE Transactions
  on}, 35\penalty0 (8):\penalty0 1872--1886, 2013{\natexlab{b}}.

\bibitem[Bruna et~al.(2014)Bruna, Sprechmann, and Lecun]{icassp14}
Bruna, J., Sprechmann, P., and Lecun, Yann.
\newblock Source separation with scattering non-negative matrix factorization.
\newblock \emph{submitted}, 2014.

\bibitem[Dong et~al.(2014)Dong, Loy, He, and Tang]{superres}
Dong, Chao, Loy, ChenChange, He, Kaiming, and Tang, Xiaoou.
\newblock Learning a deep convolutional network for image super-resolution.
\newblock In Fleet, David, Pajdla, Tomas, Schiele, Bernt, and Tuytelaars, Tinne
  (eds.), \emph{Computer Vision ? ECCV 2014}, volume 8692 of \emph{Lecture
  Notes in Computer Science}, pp.\  184--199. 2014.
\newblock \doi{10.1007/978-3-319-10593-2_13}.

\bibitem[Duan et~al.(2012)Duan, Mysore, and Smaragdis]{DuanMS12}
Duan, Z., Mysore, G.~J., and Smaragdis, P.
\newblock Online plca for real-time semi-supervised source separation.
\newblock In \emph{LVA/ICA}, pp.\  34--41, 2012.

\bibitem[F{\'e}votte(2011)]{fevotte2011majorization}
F{\'e}votte, C.
\newblock Majorization-minimization algorithm for smooth itakura-saito
  nonnegative matrix factorization.
\newblock In \emph{ICASSP}, pp.\  1980--1983. IEEE, 2011.

\bibitem[F{\'e}votte \& Idier(2011)F{\'e}votte and
  Idier]{fevotte2011algorithms}
F{\'e}votte, C. and Idier, J.
\newblock Algorithms for nonnegative matrix factorization with the
  $\beta$-divergence.
\newblock \emph{Neural Computation}, 23\penalty0 (9):\penalty0 2421--2456,
  2011.

\bibitem[F\'evotte et~al.(2013)F\'evotte, Roux, and Hershey]{icassp13a}
F\'evotte, C., Roux, J.~Le, and Hershey, J.~R.
\newblock Non-negative dynamical system with application to speech and audio.
\newblock In \emph{ICASSP}, 2013.

\bibitem[Gerchberg \& Saxton(1972)Gerchberg and Saxton]{yonina}
Gerchberg, R.~W. and Saxton, W.~Owen.
\newblock {A practical algorithm for the determination of the phase from image
  and diffraction plane pictures}.
\newblock \emph{Optik}, 35:\penalty0 237--246, 1972.

\bibitem[Gregor \& LeCun(2010)Gregor and LeCun]{LecunNN}
Gregor, K. and LeCun, Y.
\newblock Learning fast approximations of sparse coding.
\newblock In \emph{ICML}, pp.\  399--406, 2010.

\bibitem[Han et~al.(2012)Han, Mysore, and Pardo]{HanMP12}
Han, J., Mysore, G.~J., and Pardo, B.
\newblock Audio imputation using the non-negative hidden markov model.
\newblock In \emph{LVA/ICA}, pp.\  347--355, 2012.

\bibitem[H{\"a}nsler \& Schmidt(2008)H{\"a}nsler and
  Schmidt]{hansler2008speech}
H{\"a}nsler, E. and Schmidt, G.
\newblock \emph{Speech and {A}udio {P}rocessing in {A}dverse {E}nvironments}.
\newblock Springer, 2008.

\bibitem[Huang et~al.(2014{\natexlab{a}})Huang, Kim, Hasegawa-Johnson, and
  Smaragdis]{Huang_DNN_Separation_ICASSP2014}
Huang, P.-S., Kim, M., Hasegawa-Johnson, M., and Smaragdis, P.
\newblock Deep learning for monaural speech separation.
\newblock In \emph{ICASSP}, pp.\  1562--1566, 2014{\natexlab{a}}.

\bibitem[Huang et~al.(2014{\natexlab{b}})Huang, Kim, Hasegawa-Johnson, and
  Smaragdis]{huang2014singing}
Huang, Po-Sen, Kim, Minje, Hasegawa-Johnson, Mark, and Smaragdis, Paris.
\newblock Singing-voice separation from monaural recordings using deep
  recurrent neural networks.
\newblock \emph{ISMIR}, 2014{\natexlab{b}}.

\bibitem[Lee \& Seung(1999)Lee and Seung]{NMF}
Lee, D.D. and Seung, H.S.
\newblock Learning parts of objects by non-negative matrix factorization.
\newblock \emph{Nature}, 401\penalty0 (6755):\penalty0 788--791, 1999.

\bibitem[Loizou(2007)]{loizou2007speech}
Loizou, P.~C.
\newblock \emph{Speech {E}nhancement: {T}heory and {P}ractice}, volume~30.
\newblock CRC, 2007.

\bibitem[Mairal et~al.(2012)Mairal, Bach, and Ponce]{mairal2012task}
Mairal, J., Bach, F., and Ponce, J.
\newblock Task-driven dictionary learning.
\newblock \emph{Pattern Analysis and Machine Intelligence, IEEE Transactions
  on}, 34\penalty0 (4):\penalty0 791--804, 2012.

\bibitem[Mallat(1999)]{wavelettour}
Mallat, St{\'e}phane.
\newblock \emph{A wavelet tour of signal processing}.
\newblock Academic press, 1999.

\bibitem[Mallat(2010)]{eurispco}
Mallat, St{\'e}phane.
\newblock Recursive interferometric representation.
\newblock In \emph{Proc. of EUSICO conference, Denmark}, 2010.

\bibitem[Mohamed et~al.(2012)Mohamed, Hinton, and
  Penn]{mohamed2012understanding}
Mohamed, Abdel-rahman, Hinton, Geoffrey, and Penn, Gerald.
\newblock Understanding how deep belief networks perform acoustic modelling.
\newblock In \emph{Acoustics, Speech and Signal Processing (ICASSP), 2012 IEEE
  International Conference on}, pp.\  4273--4276. IEEE, 2012.

\bibitem[Mohammadiha et~al.(2013)Mohammadiha, Smaragdis, and
  Leijon]{mohammadiha2013supervised}
Mohammadiha, N., Smaragdis, P., and Leijon, A.
\newblock Supervised and unsupervised speech enhancement using nonnegative
  matrix factorization.
\newblock \emph{Audio, Speech, and Language Processing, IEEE Transactions on},
  21\penalty0 (10):\penalty0 2140--2151, 2013.

\bibitem[Mysore \& Smaragdis(2011)Mysore and Smaragdis]{MysoreS11}
Mysore, G.~J. and Smaragdis, P.
\newblock A non-negative approach to semi-supervised separation of speech from
  noise with the use of temporal dynamics.
\newblock In \emph{ICASSP}, pp.\  17--20, 2011.

\bibitem[Schmidt \& Olsson(2006)Schmidt and Olsson]{schmidt06speechseparation}
Schmidt, M.~N. and Olsson, R.~K.
\newblock Single-channel speech separation using sparse non-negative matrix
  factorization.
\newblock In \emph{INTERSPEECH}, Sep 2006.

\bibitem[Schmidt et~al.(2007)Schmidt, Larsen, and Hsiao]{schmidt07mlsp}
Schmidt, M.~N., Larsen, J., and Hsiao, F.-T.
\newblock Wind noise reduction using non-negative sparse coding.
\newblock In \emph{MLSP}, pp.\  431--436, Aug 2007.

\bibitem[Schuler et~al.(2014)Schuler, Hirsch, Harmeling, and
  Scholkopf]{deblur_mpi}
Schuler, Ch., Hirsch, M., Harmeling, S., and Scholkopf, B.
\newblock Learning to deblur.
\newblock \emph{arXiv preprint arXiv:1406.7444}, 2014.

\bibitem[Shashanka et~al.(2007)Shashanka, Raj, and
  Smaragdis]{shashanka_icassp07}
Shashanka, M. V.~S., Raj, B., and Smaragdis, P.
\newblock {Sparse Overcomplete Decomposition for Single Channel Speaker
  Separation}.
\newblock In \emph{ICASSP}, 2007.

\bibitem[Simonyan \& Zisserman(2014)Simonyan and Zisserman]{matconvnet}
Simonyan, Karen and Zisserman, Andrew.
\newblock Very deep convolutional networks for large-scale image recognition.
\newblock \emph{arXiv preprint arXiv:1409.1556}, 2014.

\bibitem[Smaragdis et~al.(2014)Smaragdis, Fevotte, Mysore, Mohammadiha, and
  Hoffman]{smaragdis2014static}
Smaragdis, P., Fevotte, C., Mysore, G, Mohammadiha, N., and Hoffman, M.
\newblock Static and dynamic source separation using nonnegative
  factorizations: A unified view.
\newblock \emph{Signal Processing Magazine, IEEE}, 31\penalty0 (3):\penalty0
  66--75, 2014.

\bibitem[Sprechmann et~al.(2013)Sprechmann, Bronstein, Bronstein, and
  Sapiro]{sprechmann2013learnable}
Sprechmann, P., Bronstein, A., Bronstein, M., and Sapiro, G.
\newblock Learnable low rank sparse models for speech denoising.
\newblock In \emph{ICASSP}, pp.\  136--140, 2013.

\bibitem[Sprechmann et~al.(2014)Sprechmann, Bronstein, and
  Sapiro]{sprechmann2014supervised}
Sprechmann, P., Bronstein, A.~M., and Sapiro, G.
\newblock Supervised non-euclidean sparse {NMF} via bilevel optimization with
  applications to speech enhancement.
\newblock In \emph{HSCMA}, pp.\  11--15. IEEE, 2014.

\bibitem[Sprechmann et~al.(2012)Sprechmann, Bronstein, and
  Sapiro]{sprechmann2012real}
Sprechmann, Pablo, Bronstein, Alexander~M, and Sapiro, Guillermo.
\newblock Real-time online singing voice separation from monaural recordings
  using robust low-rank modeling.
\newblock In \emph{ISMIR}, pp.\  67--72. Citeseer, 2012.

\bibitem[Vincent et~al.(2006)Vincent, Gribonval, and
  F{\'e}votte]{vincent2006performance}
Vincent, E., Gribonval, R., and F{\'e}votte, C.
\newblock Performance measurement in blind audio source separation.
\newblock \emph{IEEE Trans. on Audio, Speech, and Lang. Proc.}, 14\penalty0
  (4):\penalty0 1462--1469, 2006.

\bibitem[Weninger et~al.(2014{\natexlab{a}})Weninger, Le~Roux, Hershey, and
  Watanabe]{weninger2014discriminative}
Weninger, F., Le~Roux, J., Hershey, J.~R, and Watanabe, S.
\newblock Discriminative {NMF} and its application to single-channel source
  separation.
\newblock \emph{Proc. of ISCA Interspeech}, 2014{\natexlab{a}}.

\bibitem[Weninger et~al.(2014{\natexlab{b}})Weninger, {Le Roux}, Hershey, and
  Schuller]{Weninger2014GlobalSIP12}
Weninger, Felix, {Le Roux}, Jonathan, Hershey, John~R., and Schuller,
  Bj{\"o}rn.
\newblock Discriminatively trained recurrent neural networks for single-channel
  speech separation.
\newblock In \emph{Proc. IEEE GlobalSIP 2014 Symposium on Machine Learning
  Applications in Speech Processing}, 2014{\natexlab{b}}.

\bibitem[Wilson et~al.(2008)Wilson, Raj, Smaragdis, and Divakaran]{WilsonRSD08}
Wilson, K.~W., Raj, B., Smaragdis, P., and Divakaran, A.
\newblock Speech denoising using nonnegative matrix factorization with priors.
\newblock In \emph{ICASSP}, pp.\  4029--4032, 2008.

\end{thebibliography}
\bibliographystyle{iclr2015}

\end{document}